\documentclass[aps,showpacs,amsmath,superscriptaddress,floatfix,nobalancelastpage,twocolumn, prl]{revtex4-1}

\usepackage[nottoc,numbib]{tocbibind}

\usepackage[colorlinks=true, citecolor=black, linkcolor=black, urlcolor=black]{hyperref}

\usepackage[all]{hypcap}
\usepackage{amsmath}
\usepackage{enumerate,bm}
\usepackage{graphicx}
\usepackage{grffile} 
\usepackage{color}
\usepackage{esint}
\usepackage{textpos}
\usepackage[usenames,dvipsnames]{xcolor}
\usepackage{subfigure}
\usepackage{flushend}
\usepackage{tabularx}
\usepackage{amssymb}
\usepackage{float}
\usepackage{subfigure}
\usepackage{ulem} 

\usepackage{lineno} 
\usepackage{setspace}
\usepackage{verbatim}
\usepackage{ragged2e}
\usepackage{url}

\floatstyle{ruled}
\newfloat{Box}{thp}{lop}

\begin{document}

\author{Umberto~De~Giovannini*}
\email{umberto.de-giovannini@mpsd.mpg.de}
\affiliation{Max Planck Institute for the Structure and Dynamics of Matter and Center for Free Electron Laser Science, 22761 Hamburg, Germany}

\author{Hannes~H\"ubener*}
\email{hannes.huebener@mpsd.mpg.de}
\affiliation{Max Planck Institute for the Structure and Dynamics of Matter and Center for Free Electron Laser Science, 22761 Hamburg, Germany}

\author{Shunsuke~A.~Sato}
\email{ssato@ccs.tsukuba.ac.jp}
\affiliation 
{Center for Computational Sciences, University of Tsukuba, Tsukuba 305-8577, Japan}
\affiliation{Max Planck Institute for the Structure and Dynamics of Matter and Center for Free Electron Laser Science, 22761 Hamburg, Germany}

\author{Angel Rubio}
\email{angel.rubio@mpsd.mpg.de}
\affiliation{Max Planck Institute for the Structure and Dynamics of Matter and Center for Free Electron Laser Science, 22761 Hamburg, Germany}
\affiliation{Center for Computational Quantum Physics (CCQ), The Flatiron Institute, 162 Fifth avenue, New York NY 10010.}

\title{Direct measurement of electron-phonon coupling with time-resolved ARPES}

\begin{abstract}
Time and angular resolved photoelectron spectroscopy is a powerful technique to measure electron dynamics in solids. Recent advances in this technique have facilitated band and energy resolved observations of the effect that excited phonons, have on the electronic structure. Here, we show with the help of \textit{ab initio} simulations that the Fourier analysis of time-resolved measurements of solids with excited phonon modes leads, in fact, to an observation of the band- and mode-resolved electron-phonon coupling directly from the experimental data and without need for theoretical computations. 
\end{abstract}

\pacs{0.0}

\maketitle

Electron-phonon coupling is one of the fundamental interactions in solids that determines a vast range of material phenomena, ranging from thermal properties like specific heat over carrier mobility in charge transport to the critical temperature of conventional superconductivity. Several theoretical concepts have been defined in order to quantify this interaction, such as the mass enhancement factor in metals to describe band velocity renormalization. It is however not unambiguously observable~\cite{Calandra:2007,Park:2008fr} so that electron-phonon coupling remains a largely theoretical concept that is invoked to explain phenomena rather than being a direct measurable quantity, despite some success in quantifying it with inelastic scattering experiments~\cite{FERRARI200747,PhysRevB.76.035439,PhysRevLett.87.037001}. Some spectroscopic signatures in solids have been directly linked to strong electron-phonon coupling. Most prominently in photo-electron spectra electron-phonon coupling can lead to satellites~\cite{Lee:2014bw,Wang:2016de} and the mentioned kinks~\cite{Lanzara:2001kp,Mazzola:2013if} that result from (collective) excitation of phonon modes in the photo-emission process. But also optical properties are determined by electron phonon coupling through the temperature dependent renormalization of the band gap~\cite{Cardona:2005ho,Giustino:2010ek} and phonon assisted absorption~\cite{Noffsinger:2012hf}, as well as excited state lifetimes and carrier relaxation~\cite{Bernardi:2014hi,Gierz:2015jy,Gierz:2015gn,Waldecker:2017kq}, which can also be used to extract electron-phonon matrix elements if the scattering path is well defined~\cite{10.1126/science.aaw1662}.

Recent experimental advances in time and angular resolved photoelectron spectroscopy (tr-ARPES), however, have shown that a more direct observation of electron phonon coupling is possible~\cite{hein:2019,suzuki:2020}. These works show that with sufficiently clean data it is possible to extract some electron-phonon coupling properties by Fourier analysis of tr-ARPES data, in a method called frequency-domain ARPES (FD-ARPES) by some authors~\cite{suzuki:2020}. While these works are pioneering the use of tr-ARPES to observe the effect of phonons on the electronic structure in solids the underlying details of electron-phonon coupling remain unknown. 

The effect of lattice distortions on the electronic bandstructure is often described in terms of so called "frozen phonon" bands, the instantaneous electronic structure for a given lattice configuration. The assumption is that this reflects the dynamics of the bandstructure and hence can be used to interpret tr-ARPES measurements. However, its observation requires spectrally sharp probe pulses with durations well below the phonon cycle, a bandwidth smaller than the phonon-induced change in energy and for sufficiently sharp intrinsic lineshapes. These conditions can be met for slow phonons and with accurate equipment~\cite{Gerber:2017bm}, but even then the results are confined to those regions in the Brillouin zone (BZ) where the phonon induced change in band energy is large, in other words where the electron-phonon coupling is high.   

In this letter we show that FD-ARPES is an experimental technique that can directly observe the electron-phonon coupling matrix elements~\cite{Giustino:2017ge} of single electronic bands with momentum space and mode resolution, thus elevating electron-phonon coupling to an observable of photoelectron spectroscopy. It has far reaching potential for the investigation of fundamental processes in solids including excited state dynamics, phase transitions and the characterization of novel non-equilibrium phases. FD-ARPES requires tr-ARPES probes that are fast enough to resolve the phonon dynamics, i.e. have a time resolution that is at least of the same order of magnitude as the phonon frequency. However, the faster a probe pulse is, the broader is its resulting ARPES spectrum. Therefore, it will be helpful for the following discussion to consider the ratio between the pulse dependent spectral linewidth $\sigma$ and the phonon-induced variation in band energy $\Delta$ (defined below). When this \textit{probe-phonon spectral resolution} is $\Delta<<\sigma$ a time-resolved observation of $\Delta$ is not possible and one has to consider FD-ARPES. We will see that even if $\Delta>\sigma$ a careful analysis of FD-ARPES is required. We therefore consider these two regimes in this letter, where first the phonon induced variation in band energy is small compared with the spectral resolution and second where both are comparable or larger. FD-ARPES requires coherent phonon excitation, but does not rely on carrier relaxation, which makes it well suited to probe the coupling of a wide range of electronic bands to the same phonon mode.
Our results are based on first-principles time-dependent density functional theory (TDDFT) calculations, however, the discussed method of analysis of photoelectron spectra can be directly applied to experimental data, without input from calculations. We demonstrate this method by using the example of graphene and a single phonon mode, but it is general and can be applied to any material.   

The main findings of this letter can be summarized as follows: (i) FD-ARPES is a technique to observe electron-phonon coupling with momentum and energy resolution. (ii) The lineshape of the FD-ARPES signal contains information about interband electron-phonon coupling. (iii) Also for the case where $\Delta >> \sigma$, i.e. well resolvable band changes in tr-ARPES, the  FD-ARPES provides appropriate tool to interpret time-dependent bandstructures.

To analyze the dynamics underlying the FD-ARPES measurement we consider a Fermi's Golden Rule expression for the detected ARPES intensity originating from a single band as the product of a photo-electron matrix element and a spectral lineshape $F$
\begin{equation}\label{eq:arpes}
I_{i{\bf k}}(E_{\rm kin}) = |\langle f_{\bf p} | {\bf A}\cdot \hat{\bf p} |\psi_{i{\bf k}}\rangle|^2 F(\Omega+\epsilon_{i\mathbf{k}}+E_{\rm kin})
\end{equation}
where $E_{\rm kin} =p^2/2$ is the kinetic energy of the photoelectron, $f_{\bf p}$ is a final state with momentum $\mathbf{p}$,   $\psi_{i\mathbf{k}}$ and $\epsilon_{i\mathbf{k}}$ are the (initial) band orbitals and energies and $\Omega$ is the energy of the probe laser. The spectral lineshape $F$ is determined by the power spectrum of the probe pulse. We stress, that the above expression is not used here to compute the ARPES spectra, but is considered to understand the FD-ARPES concept. For the same reason, we also consider only a single coherent phonon mode, while in actual pump-probe experiments whole range of phonons can be excited simultaneously with, for instance, impulsive Raman techniques, and analyzed with FD-ARPES. A coherent phonon mode with the frequency $\omega_0$ induces a variation in the band energy and orbital that can be parametrized by the instantaneous lattice distortion $ u(\tau)=u_0\sin(\omega_0 \tau)$ if the electron-phonon coupling is adiabatic. The band properties then read to linear order in the displacement, i.e. if $u_0$ is sufficiently small,
\begin{align}
    \epsilon_{i\mathbf{k}}[ u(\tau)] & =  \epsilon_{i\mathbf{k}} + \Delta_{i{\bf k}} \sin{(\omega_0\tau)} \\
   | \psi_{i\mathbf{k}}[ u(\tau)]\rangle & = |\psi_{i\mathbf{k}}\rangle + |\delta \psi_{i\mathbf{k}} \rangle  \sin{(\omega_0\tau)}
\end{align}
where $\Delta_{i{\bf k}} = u_0\langle \psi_{i\mathbf{k}} |\delta V |\psi_{i\mathbf{k}} \rangle = u_0g_{ii}({\bf k})$ depends on the diagonal electron-phonon coupling matrix element, i.e. the matrix element of the deformation potential $\delta V$ that results from the change in the lattice configuration~\cite{Giustino:2017ge}. The term $|\delta \psi_{i\mathbf{k}} \rangle = u_0 \sum_{i\neq j} \frac{g_{ij}({\bf k})}{ \epsilon_{i\mathbf{k}}- \epsilon_{j\mathbf{k}}} | \psi_{j\mathbf{k}}\rangle$ depends on the interband electron-phonon coupling. These time-dependent band parameters confer their time dependence to the ARPES spectrum via Eq.~(\ref{eq:arpes}) giving the time-dependent adiabatic ARPES signal as
\begin{align}
    I_{i{\bf k}}(E_{\rm kin},\tau) = & |\langle f_{\bf p} | {\bf A}\cdot \hat{\bf p} |\psi_{i{\bf k}}[u(\tau)]\rangle|^2 \nonumber\\
    &\times F(\Omega+\epsilon_{i\mathbf{k}}[u(\tau)]+E_{\rm kin}) .
\end{align}
FD-ARPES is obtained from such a time-dependent signal as the Fourier transform at a fixed frequency for each $E_{\rm kin}$ and ${\bf k}$. 

\begin{figure}
\centering
\includegraphics[width=1\columnwidth]{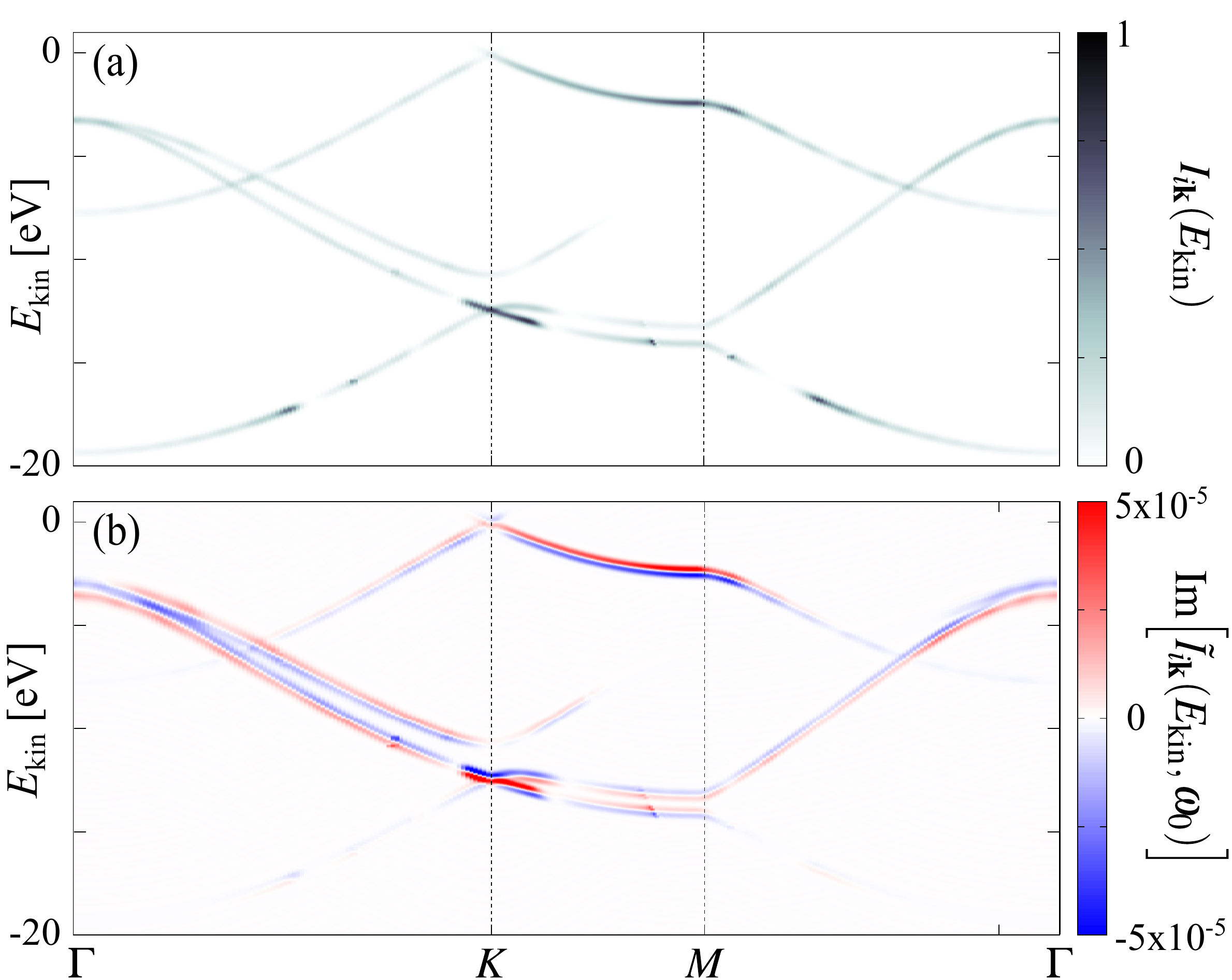}
   \caption{(a) Equilibrium ARPES spectrum of graphene computed with TDDFT (see text) along a path of high symmetry points in the 2nd~BZ. (b) FD-ARPES at the frequency of the $E_{2g}$ mode, showing quantitatively which bands couple strongly to this mode. The phase of the complex FD-APRES, which is determined by the phase of the coherent phonon mode, leads to purely imaginary signal in this example calculation.}
\end{figure}

To demonstrate this technique and its analysis, we consider computed tr-ARPES spectra of graphene with a single coherent $E_{2g}$ phonon mode using TDDFT~\cite{runge_density-functional_1984} together with Ehrenfest molecular dynamics. The $E_{2g}$ optical mode of graphene is one of the fastest phonon modes in materials with a frequency of $\sim$~48 THz, corresponding to a cycle time of $\sim$~20 fs, and therefore presents a challenge to directly resolve the phonon induced dynamics with tr-ARPES, which makes it a good candidate to showcase this method. The electron-phonon coupling for this mode is most pronounced for the $\sigma$ bands near the $\Gamma$-point of the electronic BZ and comparatively weak for other other bands~\cite{Mazzola:2017ih}. The DFT electronic groundstate, time propagation and photoelectron spectra were obtained with the octopus code~\cite{TancogneDejean:2020ek} using a 12$\times$12$\times$1 sampling of the BZ, a realspace sampling with a spacing of 0.36~Bohr atomic units and the (adiabatic) local density approximation. The time-resolved photoelectron spectra for pulses with FWHM of 10~fs and carrier energy of 80~eV were computed using the t-SURFF technique~\cite{Tao:2012ev} implemented in octopus~\cite{deGiovannini:2016bb,deGiovannini:2016cb,schuler2020}, with the flux surface being located 30~Bohr atomic units above the graphene sheet and the resulting equilibrium ARPES spectrum is shown in Fig.~1~(a). The $E_{2g}$ phonon mode can be included by setting an initial velocity on the ions along the eigenmode displacement direction and by then  co-propagating the Ehrenfest dynamics together with the TDDFT~\cite{Andrade:2009ga,Shin:2018}. Here, we simplified this approach by computing the trajectories for the ions only once with the TDDFT+Ehrenfest and used it for all tr-ARPES calculations, without any loss in numerical accuracy, because the ARPES probe intensity is weak enough to not affect the phonon behaviour.

\begin{figure}[t]
\centering
\includegraphics[width=1\columnwidth]{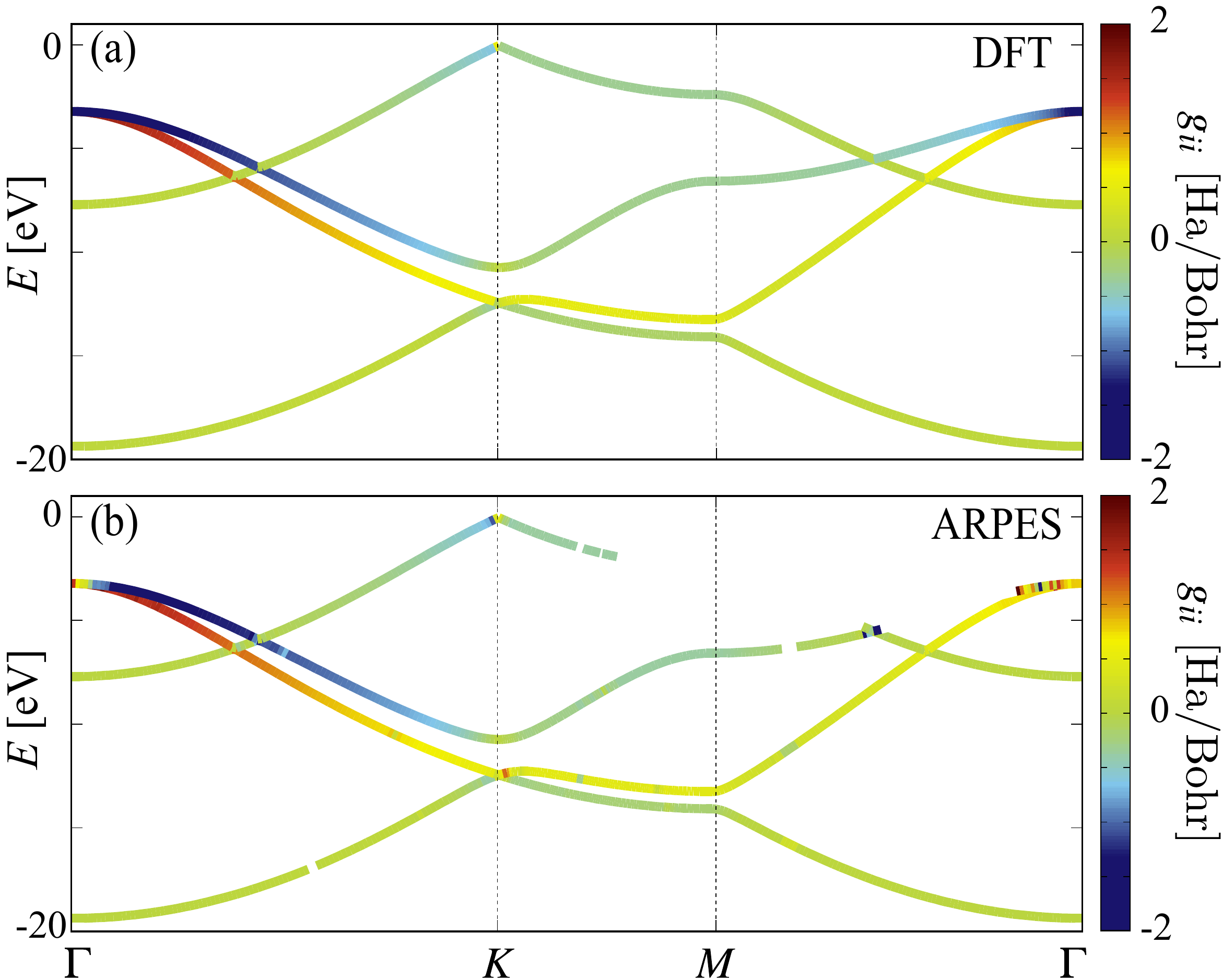}
   \caption{(a) The theoretical electron phonon-coupling matrix element $g_{ii}({\bf k})$ (see text) obtained from static density functional calculations. (b) The same quantity obtained by analyzing computed FD-ARPES spectrum in the linear (small displacement) regime, c.f Eq.~(6). Missing points have been excluded because of too weak equilibrium ARPES intensity (c.f. Fig.~1).}
\end{figure}

The tr-ARPES spectra $I_{\bf k}(E_{\rm kin},\tau)$ were computed for $N_\tau=9$ time delays $\tau_i$ during the phonon cycle and subsequently the Fourier transform for the phonon frequency $\omega_0$ was obtained as $\tilde{I}_{\bf k}(E_{\rm kin},\omega_0) = \frac{1}{N_\tau}\sum_j \exp{(-i \omega_0 \tau_j)} I_{\bf k}(E_{\rm kin},\tau_j)$, for each $k$-point along a $\Gamma-M-K$ path in the 2nd~BZ and for each point sampling the spectral energy $E$, shown in Fig.~1~(b). An experimental FD-ARPES measurement could of course take more time-samples and over a longer time to compute the Fourier transform at all frequencies, but the basic data processing is the same as in this illustration. Such treatment would yield information on all phonon modes and without bias of a given frequency. The choice of a path in the 2nd~BZ was taken, because here the $\sigma$-bands at the $\Gamma$-point have larger photoelectron matrix elements, leading to better resolved spectra~\cite{Mazzola:2017ih,Hubener:2018id}. In the resulting FD-ARPES spectrum, shown in Fig.~1~(b), one can already qualitatively observe from the intensity which bands are strongly coupled to the phonon mode across the BZ and, for instance, by analyzing the phonon-pump fluence dependence of these signals one can obtain a qualitative measure of electron-phonon coupling. In the remainder of this letter however, we will analyse how detailed quantitative information can be obtained from the lineshape of the FD-ARPES signal. 

\textit{The $\Delta << \sigma$ regime. } The dependence of the FD-ARPES on the phonon displacement $u$ is highly non-linear so even a small variation in the band properties can result in strong changes in the ARPES spectrum. We therefore first consider those cases where the lattice distortion of the phonon induces only small changes in the spectrum. We simplify the notation by dropping the $i{\bf k}$ index in the right hand side of the following equations. 
To linear order in the displacement we then have for a single band the ARPES intensity $I_{i{\bf k}}(E_{\rm kin},\tau) \approx I^{(0)}_{i{\bf k}}(E_{\rm kin}) + I_{i{\bf k}}^{(1)}(E_{\rm kin},\tau)$ with 
\begin{align}
    I_{i{\bf k}}^{(1)}(E_{\rm kin},\tau) 
    =& M  F'(E) \Delta \sin{(\omega_0\tau)}\nonumber\\
     &+ \left.\frac{\partial M[u(\tau)]}{\partial u (\tau)}\right|_{u=0}u_0\sin{(\omega_0\tau)} F(E) 
\end{align}
where we defined $E=\Omega+\epsilon_{i\mathbf{k}}+E_{\rm kin}$, $M=|\langle f_{\bf p} | {\bf A}\cdot \hat{\bf p} |\psi_{i{\bf k}}\rangle|^2$  and $F'(E) =\frac{\partial F }{\partial E}$ is the derivative of the spectral lineshape. The Fourier series expansion over $\tau$ of the time-resolved ARPES signal at the frequency $\omega_0$ reads under this condition
\begin{align}
    \tilde{I}_{i{\bf k}}^{(1)}(E_{\rm kin}, \omega_0)= \frac{i}{2} M \Delta F'(E)
     + \frac{i}{2}\left.\frac{\partial M[u(\tau)]}{\partial u (\tau)}\right|_{u=0}u_0 F(E) .
\end{align}
This illustrates important properties of the FD-ARPES signal. The first term is directly proportional to the diagonal electron-phonon coupling and appears with the derivative of the equilibrium lineshape, while the second term, depending on off-diagonal electron phonon coupling, contributes with the same line-shape as the equilibrium. Since, here we made no further assumption on the lineshape, this meas  FD-ARPES spectra can be analysed by comparison with the equilibrium lineshape and yield information on different electron-phonon coupling matrix elements with momentum and band resolution, even in regions where the FD-ARPES signal depends only weakly on the electron-phonon coupling.

To demonstrate how this information is directly obtained from experimental data we computed the FD-ARPES spectrum for graphene with the $E_{2g}$ mode with amplitude $u_0=0.001$~a$_{C-C}$, where a$_{C-C}$ is the carbon bondlength in graphene. The resulting $\tilde{I}$ scales linearly with small lattice displacements, so that Eq.~(6) can be applied. The first term in Eq.~(6) is directly proportional to the energy derivative of the \textit{equilibrium} ARPES spectrum $I^{(0)}$, assuming that the matrix element does not depend on energy within the range of $F$, and can be readily computed by finite differences from such data, while the second term contains directly the equilibrium lineshape. By fitting the sum ${\rm Im}\Big[ \tilde{I}^{(1)}_{i \bf k}(\omega_0)\Big] = \alpha \partial  I^{(0)}_{i \bf k}/ \partial E + \beta I^{(0)}_{i \bf k}$ around each equilibrium band one directly obtains $\alpha = g_{ii}({\bf k}) u_0/2$ and $\beta=\partial M/\partial u (\tau) u_0/2$. The result of such fitting is shown in Fig.~2~(b) and the $\beta$ term did not contribute above numerical accuracy, so that the fitting directly gives the electron-phonon coupling of each band in units $u_0$. However, in situations where the $\beta$ term is significant it contains information of the inter band electron-phonon coupling via Eq.~(3), albeit in a mixture with the photo-electron matrix element $M$ that can only be disentangled with further assumptions or knowledge about $M$. The present computational demonstration of the FD-ARPES technique, allows to independently compute the electron-phonon coupling matrix element $g_{ii}$ and the comparison shown in Fig.~2 reveals the high fidelity of this method throughout the BZ. Missing points in the results only occur when either the equilibrium ARPES signal is too weak or the underlying $\Delta$ is too small. 

\textit{The $\Delta >\sigma$ regime. } When the effect of the electron-phonon coupling on the tr-ARPES spectrum leads to a non-linear change in the FD-ARPES one has to consider the Fourier transform of Eq.~(4) instead of the expansion. For the phonon frequency $\omega_0$ this reads
\begin{align}\nonumber
    \tilde{I}_{i{\bf k}}(E_{\rm kin},\omega_0) =& \frac{1}{2\Delta} M  \tilde{J}_1\bigg(\frac{E}{\Delta}\bigg)* F(E) +\\
      &\frac{i}{2\Delta}  \delta M  \left[\tilde{J}_0\bigg(\frac{E}{\Delta}\bigg)+\frac{3}{2}\tilde{J}_2\bigg(\frac{E}{\Delta}\bigg)\right] * F(E)
\end{align}
where $\delta M=Re[\langle \psi_{i{\bf k}}|{\bf A}\cdot \hat{\bf p}|f_{\bf p}\rangle \langle f_{\bf p}|{\bf A}\cdot \hat{\bf p}| \delta\psi_{i{\bf k}}\rangle ]$, $\tilde{J}_i$ are Fourier transforms of Bessel functions of the first kind, and $*$ indicates the convolution product. This expression again allows to disentangle contributions of different electron-phonon couplings, because the first term again only depends on the diagonal electron-phonon coupling matrix element via $\Delta$, while the second term depends on the inter-band coupling via $\delta M$. Assuming that the equilibrium spectral lineshape $F$ has an even symmetry, the first term gives an odd FD-ARPES lineshape and the second term a purely even contribution. This symmetry property presents another straightforward opportunity to process the FD-ARPES data: By taking the Fourier transform of the FD-ARPES at an isolated band the even and odd contributions to the lineshape become real and imaginary components in the time domain and the convolutions in Eq.~(7) become direct products. Then, the first term of Eq.~(7) is particularly simple, because it is a product of the first Bessel function and the inverse Fourier transform of the equilibrium ARPES signal $I^{(0)}=MF$, Eq.~(1), and we can fit ${\rm Im}\left[\mathcal{F}^{-1}[\tilde{I}_{i\bf k}](t)\right]=\alpha t {\rm Re}\left[\mathcal{F}^{-1}[I^{(0)}_{i\bf k}](t)\right]$ around each isolated band. Here, we have expanded the Bessel function around small time arguments $J_1(\Delta t) \approx \frac{\Delta}{2}t$ so that $g_{ii}({\bf k}) = 2\alpha/u_0$ and the fitting has to match only for a small interval in the time domain. By processing tr-ARPES data of graphene obtained with TDDFT and a lattice displacement of $u_0=0.01$~a$_{C-C}$ in this way, we obtain the result shown in Fig.~3. This method only applies if bands are sufficiently isolated to obtain a well defined signal for the inverse Fourier transform. It does, however, allow to quantify the interband contribution, via the second term, which contributes to ${\rm Re}\left[\mathcal{F}^{-1}[\tilde{I}_{i\bf k}](t)\right]$ which is proportional to  $\delta M$. We stress that Eq.~(7), being the full expression for FD-ARPES, also applies to the small $\Delta << \sigma$ regime. 
\begin{figure}
\centering
\includegraphics[width=1\columnwidth]{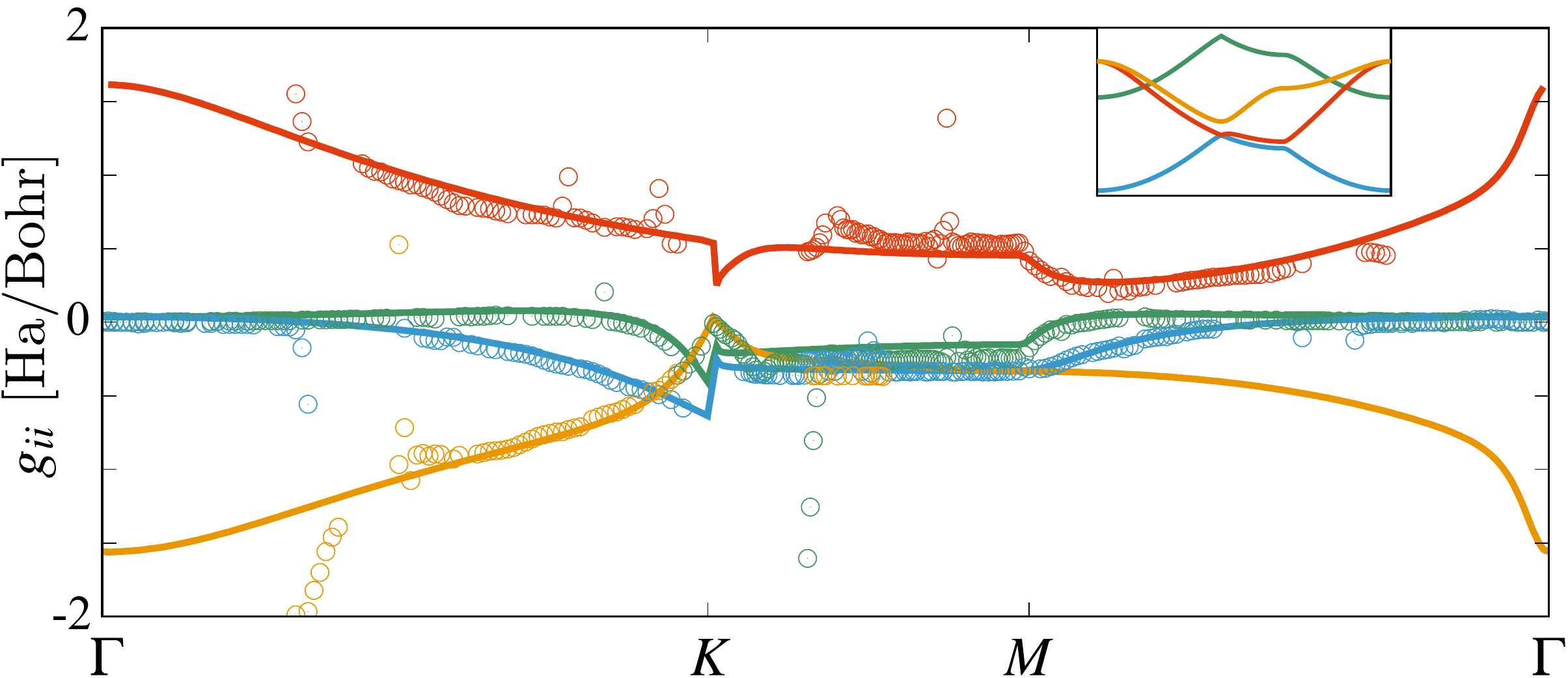}
   \caption{Electron phonon matrix element  $g_{ii}({\bf k})$ (see text) obtained from static density functional calculations (solid lines), same as in Fig.~2~(a), and from analyzing FD-ARPES data with $u_0=0.01$~a$_{C-C}$ (circles). Points close to band crossings have been excluded, because the method requires isolated bands. Color coding is the same as for the band structure in the inset.  }
\end{figure}

The expression for the full FD-ARPES signal, Eq.~(7) also gives relevant information for the case where the electron-phonon induced variation is large compared to the width of the tr-ARPES spectral function, i.e. when the phonon-probe spectral resolution $\Delta>>\sigma$. In this case only the Fourier transforms of the Bessel functions in Eq.~(7) contribute to the FD-ARPES signal and they provide sharp signals spaced by $\Delta$. This reflects the intuitive picture of a band oscillating in real time, where FD-ARPES resolves the frequency and amplitude of this oscillation. Such an oscillation that is large enough to be resolved by ARPES linewidth can be observed directly by the tr-ARPES signal~\cite{Gerber:2017bm}, however here we argue that the analysis of the time-resolved signal amounts to performing FD-ARPES and represents just one limiting case of this technique that can be applied across all ranges of phonon induced variations in the spectrum.

We have demosntrated how FD-ARPES gives direct access to the band resolved electron-phonon coupling matrix element even for small coupling and fast phonons by analysing the experimental data without requiring further input from theory . The three regimes of different phonon induced changes in the tr-ARPES spectrum that we considered above do not necessarily occur in separate experiments, but all three might be observed across the BZ in the same FD-ARPES spectrum, because $\Delta$ can vary considerably for different bands and different $k$-points. For this reason we chose to present these three cases, as they most likely have to be considered all at once when interpreting experimental data. 

In this letter we considered the linear expansion of the electronic band energy and orbitals with lattice perturbation, however FD-ARPES also accesses higher order effects that would oscillate at other frequencies. Finally we stress again that the analysis presented here can be applied over a wide frequency range at once, giving access to the effect of all excited phonons. Especially it allows to track the dynamical renormalization of phonon frequencies, that might occur across phase transitions or as the result of novel driven phases.

\section{Acknowledgements}
We acknowledge financial support from the European Research Council(ERC-2015-AdG-694097). The Flatiron Institute is a division of the Simons Foundation.

UDG and HH contributed equally to the work.

\end{document}